\RequirePackage{amsmath}
\documentclass{llncs}

\usepackage[pdftex]{graphicx}
\usepackage{wrapfig}
\DeclareGraphicsExtensions{.pdf,.jpeg,.png}
\usepackage{pgf}

\usepackage{ amssymb }
\usepackage{floatrow}

\newcommand{\polsat}{\textsc{polsat}}
\newcommand{\pltl}{\textsc{pltl}}
\newcommand{\trp}{\textsc{trp++}}
\newcommand{\aalta}{\textsc{aalta}}
\newcommand{\nusmv}{\textsc{NuSMV}}
\newcommand{\nusmvi}{\textsc{NuSMV-invar}}
\newcommand{\nusmvn}{\textsc{NuSMV-noinvar}}

\newfloatcommand{capbtabbox}{table}[][\FBwidth]

\pgfdeclareimage[interpolate=true,width=155pt]{i1}{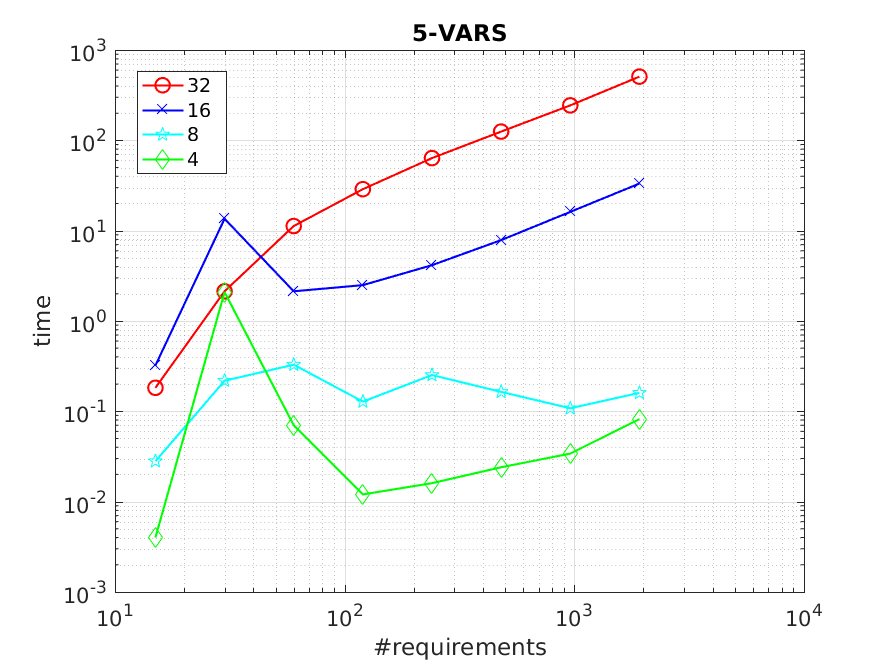}
\pgfdeclareimage[interpolate=true,width=155pt]{i2}{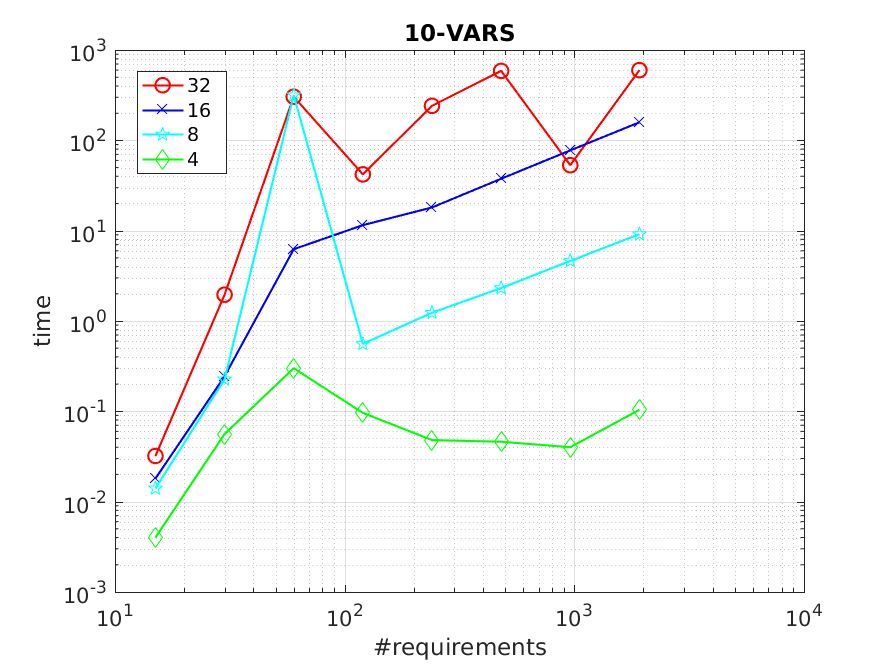}
\pgfdeclareimage[interpolate=true,width=155pt]{i3}{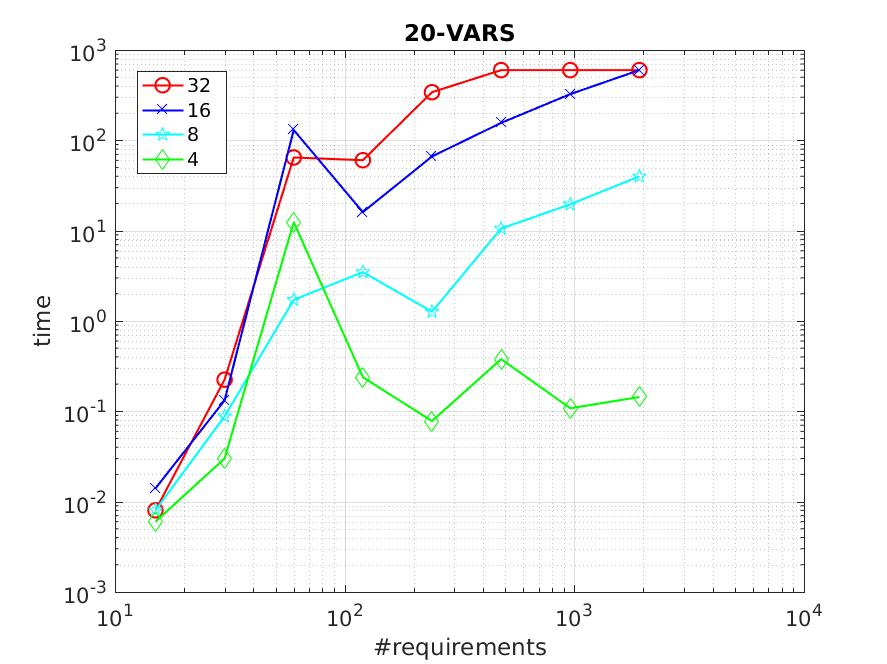}
\pgfdeclareimage[interpolate=true,width=155pt]{i4}{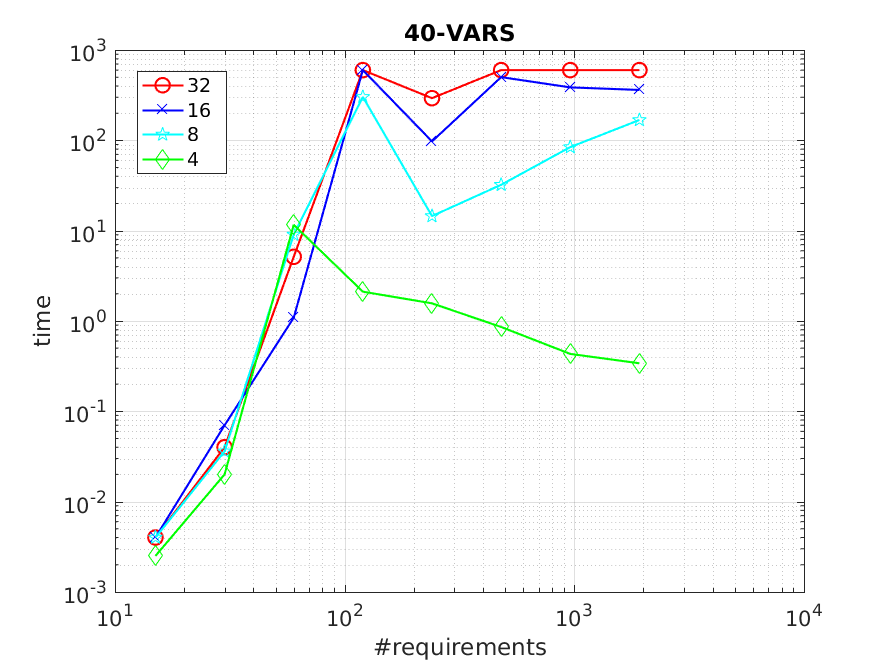}
\pgfdeclareimage[interpolate=true,width=155pt]{i5}{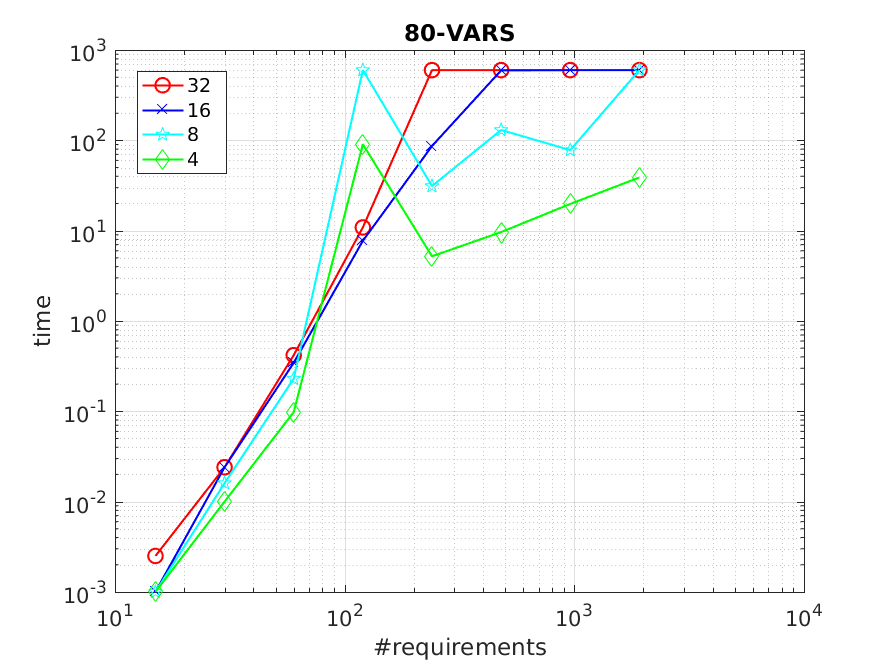}
\pgfdeclareimage[interpolate=true,width=155pt]{i6}{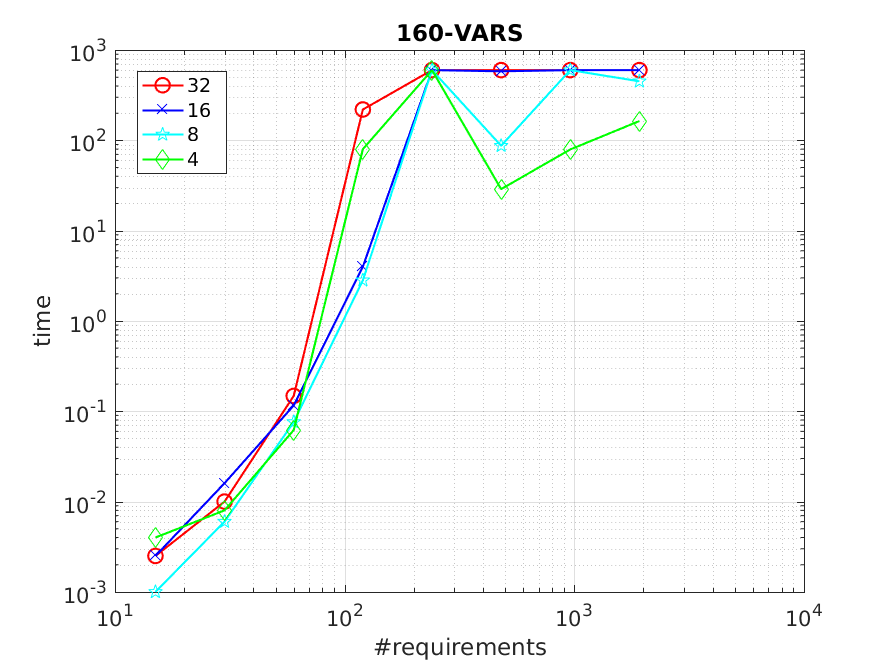}
\pgfdeclareimage[interpolate=true,width=155pt]{i7}{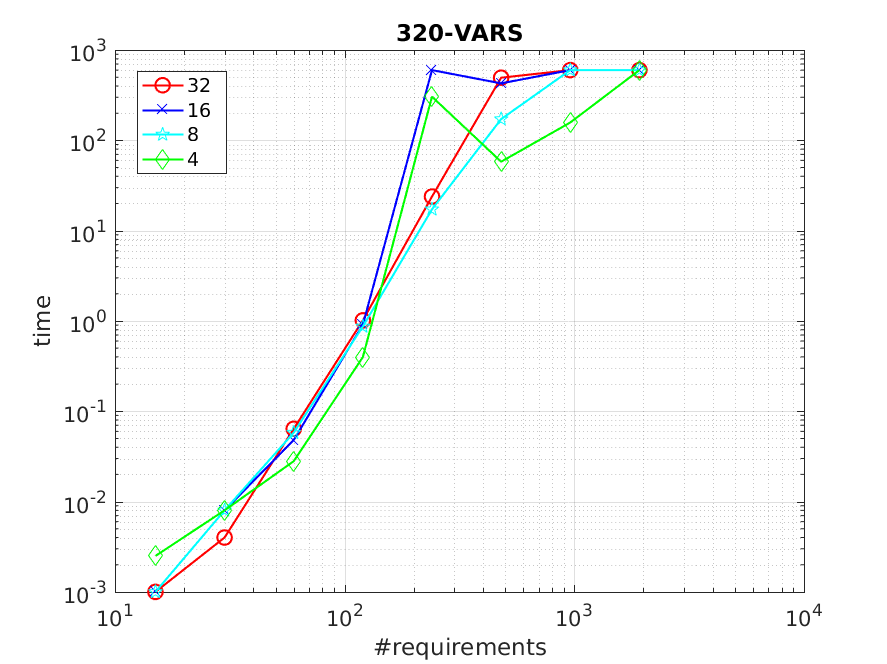}
\pgfdeclareimage[interpolate=true,width=155pt]{i8}{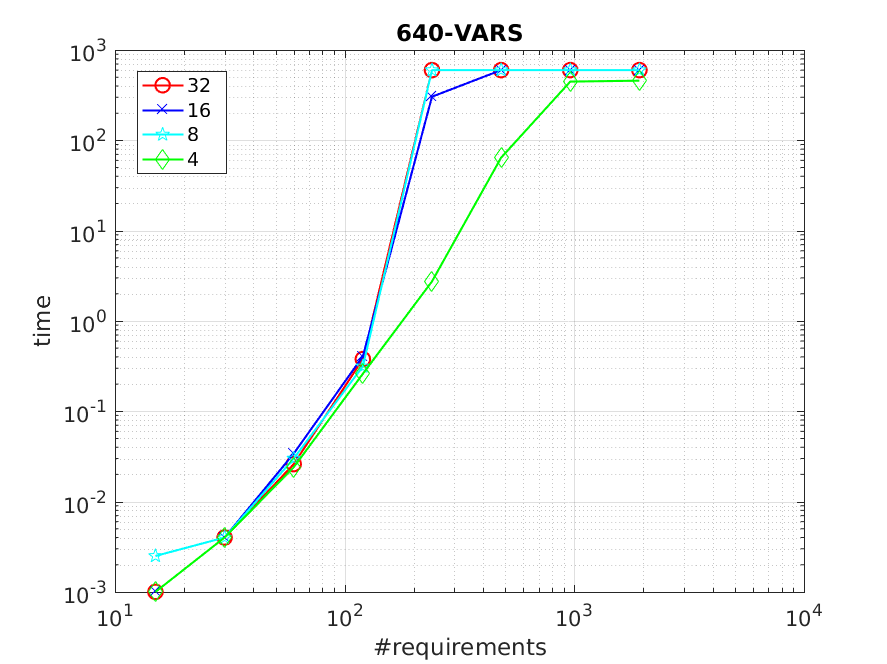}

\pgfdeclareimage[interpolate=true,width=12cm]{snl}{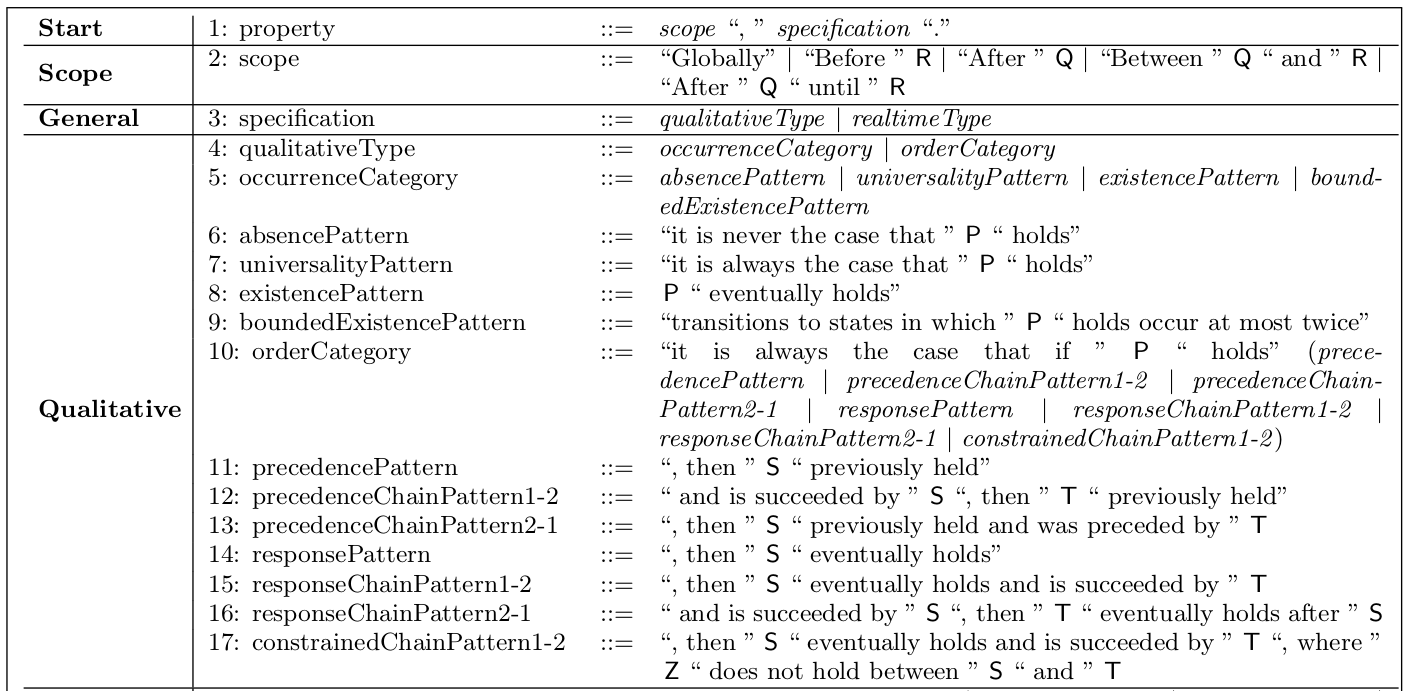}
\pgfdeclareimage[interpolate=true,width=12cm]{framework}{img/framework.png}

\begin{document}
\title{Consistency of Property Specification Patterns\\
with Boolean and Constrained Numerical Signals}

\author{Massimo Narizzano\inst{1} \and Luca Pulina\inst{2} \and
  Armando Tacchella\inst{1} \and Simone Vuotto\inst{1,2}}
\institute{DIBRIS, University of Genoa, Viale Causa 13, 16145
  Genova\\ \email{massimo.narizzano@unige.it,
    armando.tacchella@unige.it}\\ \and
  POLCOMING, University of Sassari, Viale Mancini 5, 07100
  Sassari\\ \email{lpulina@uniss.it}, \email{svuotto@uniss.it}}

\maketitle
\begin{abstract}
  Property Specification Patterns (PSPs) have been proposed
  to solve recurring specification needs, to ease the 
  formalization of requirements, and enable automated verification thereof.
  In this paper, we extend PSPs 
  by considering Boolean as well as atomic assertions from a
  constraint system. 
  This extension enables us to reason about functional
  requirements which would not be captured by basic PSPs.
  We contribute an encoding from constrained PSPs to LTL
  formulae, and we show experimental results demonstrating that our
  approach scales on requirements of realistic size generated using an
  artificial probabilistic model. Finally, we show that our extension
  enables us to prove (in)consistency of requirements about an embedded
  controller for a robotic manipulator.


\end{abstract}

\section{Introduction}
In the context of safety- and security-critical cyber-physical systems
(CPSs), checking  the consistency of functional requirements is an
indisputable, yet challenging task. Requirements written in natural
language call for time-consuming and error-prone manual reviews,
whereas enabling automated consistency verification often requires
overburdening formalizations. Given the increasing pervasiveness of
CPSs, their stringent time-to-market and product budget constraints,
practical solutions to enable automated verification of requirements
are in order, and 
Property Specification Patterns (PSPs)~\cite{dwyer1999} offer a viable
path towards this target. 
PSPs are a collection of parameterizable, high-level,
formalism-independent specification abstractions, originally developed
to capture recurring solutions to the needs of requirement engineering.
Each pattern can be directly encoded in a formal specification
language, such as linear time temporal logic (LTL)~\cite{pnueli1992},
computational tree logic (CTL)~\cite{clarke1986}, or graphical
interval logic (GIL)~\cite{dillon1994}.
Because of their features, PSPs may ease the burden of formalizing
requirements, yet enable their verification using current
state-of-the-art automated reasoning tools --- see,
e.g.,~\cite{li2013,li2013b,schwendimann1998,cimatti2002,hustadt2003}. 

The original formulation of PSPs caters for temporal structure over
Boolean variables. However, for most practical applications, such
expressiveness is too restricted. This is the case of the embedded
controller for robotic manipulators that is under development in the
context of the EU project CERBERO\footnote{Cross-layer modEl-based
  fRamework for multi-oBjective dEsign of Reconfigurable systems in
  unceRtain hybRid envirOnments ---
  \url{http://www.cerbero-h2020.eu/}} and provides the main motivation
for this work. As an example, consider the following statement:
``\textit{The angle of joint1 shall never be greater than 170
  degrees}''. This requirement imposes a safety threshold related to
some joint of the manipulator (\textit{joint1}) with respect to
physically-realizable poses, yet it cannot be expressed as a PSP
unless we add atomic assertions from a constraint system
$\mathcal{D}$.  We call Constraint PSP, or PSP($\mathcal{D}$) for
short, a pattern which has the same structure of a PSP, but contains
atomic propositions from $\mathcal{D}$. For instance, using
PSP($\mathbb{R},<,=$) we can rewrite the above requirement as an
\textit{universality} pattern: ``\textsl{Globally, it is always the
  case that $\theta_1 < 170$ holds}'', where $\theta_1$ is the
numerical signal (variable) for the angle of \textit{joint1}.  In
principle, automated reasoning about Constraint PSPs can be performed
in Constraint Linear Temporal Logic, i.e., LTL extended with atomic
assertions from a constraint system~\cite{demri2002}: in our example
above, the encoding would be simply $\Box(\theta_1 < 170)$.
Unfortunately, this approach does not always lend itself to a
practical solution, because Constraint Linear Temporal Logic
is undecidable in
general~\cite{comon2000}. Restrictions on $\mathcal{D}$ may restore
decidability~\cite{demri2002}, but they introduce limitations
in the expressiveness of the corresponding PSPs.

In this paper, we propose a solution which ensures that automated
verification of requirements is feasible, yet enables PSPs mixing
both Boolean variables and (constrained) numerical signals. Our approach
enables us to capture many specifications of practical interest,
and to pick a verification procedure from the relatively
large pool of automated reasoning systems currently available for LTL. 
In particular, we restrict our attention to a constraint systems of the
form ($\mathbb{R}$,$<$,$=$), and atomic propositions of the form
$x < C$ or $x = C$, where $x \in \mathbb{R}$ is a variable
and $C \in \mathbb{R}$ is a constant value. In the following,
we write $\mathcal{D}_C$ to denote such restriction. Our contribution
can be summarized as follows:
\begin{itemize}
\item We extended basic PSPs over the constraint system
  $\mathcal{D}_C$, and we provided an encoding from any
  PSP($\mathcal{D}_C$) into a corresponding LTL formula.
\item We implemented a generator of artificial requirements expressed
  as PSPs($\mathcal{D}_C$); the generator has a number of parameters,
  and it uses a probability model to choose
  the specific pattern to emit.
\item Using our generator, we ran an extensive experimental evaluation
  aimed at understanding $(i)$ which automated reasoning tool is best
  at handling set of requirements as PSPs($\mathcal{D}_C$), and $(ii)$
  whether our approach is scalable.
\item Finally, we analyzed the requirements of the aforementioned
  embedded controller, experimenting also with the addition of faulty
  ones.
\end{itemize}
The consistency of requirements written in PSP($\mathcal{D}_C$) is
carried out using tools and techniques available in the
literature~\cite{rozier2010,rozier2011,panda,li2013b}. With those, 
we demonstrate the scalability of our approach by checking the
consistency of up to 1920 requirements, featuring 160 variables and
domains of size 8 within less than 500 CPU seconds.
A total of 75 requirements about the embedded controller for the
CERBERO project is checked in a matter of seconds, even without
resorting to the best tool among those we consider.

The rest of the paper is organized as
follows. Section~\ref{sec:background} contains some basic concepts on
LTL, PSPs and some related work. In Section~\ref{sec:consistency} we
present the extension of basic PSPs over $\mathcal{D}_C$ and the
related encoding to LTL. In Sections~\ref{sec:random}
and~\ref{sec:robot} we report the results of the experimental analysis
concerning the scalability and the case study on the embedded
controller, respectively. We conclude the paper in
Section~\ref{sec:concl} with some final remarks.

\section{Background and Related Work}
\label{sec:background}



\paragraph{LTL syntax and semantics.} Linear temporal logic
(LTL)~\cite{pnueli1977} formulae
are built on a finite set $Prop$ of atomic propositions
as follows:
\begin{center}
  $\phi =$ $\bot$ $\mid$ $\top$ $\mid$ $p$ $\mid$ $\neg \phi_1$
  $\mid$ $\phi_1\vee\phi_2$ $\mid$ $\mathcal{X}
  \phi_1$ $\mid$ $\phi_1 \mathcal{U} \phi_2$ 
\end{center}
where $p \in Prop$, $\phi, \phi_1, \phi_2$ are LTL
formulae,  $\mathcal{X}$ is the ``next'' operator and $\mathcal{U}$ is
the ``until'' operator. An LTL formula is interpreted over 
a \emph{computation}, i.e., a function $\pi: \mathbb{N} \rightarrow
2^{Prop}$ which assigns truth values to the elements of $Prop$ at
each time instant (natural number). For a 
computation $\pi$ and a point $i \in \mathbb{N}$:
\begin{itemize}
\item $\pi, i \not\models \bot$ and $\pi, i \models \top$
\item $\pi, i \models p$ for $p \in Prop$ iff $p \in \pi(i)$
\item $\pi, i \models \neg \alpha$ iff $\pi, i \not\models \alpha$
\item $\pi, i \models (\alpha \vee \beta)$ iff
    $\pi, i \models \alpha$ or $\pi, i \models \beta$
\item $\pi, i \models \mathcal{X} \alpha$ iff $\pi, i+1 \models \alpha$
\item $\pi, i \models \alpha \mathcal{U} \beta$ iff for some $j\geq i$, we have
  $\pi, j \models \beta$ and for all $k$, $i \leq k < j$ we have
  $\pi, k \models \alpha$
\end{itemize}
We say that $\pi$ \emph{satisfies} a formula $\phi$, denoted $\pi \models
\phi$, iff $\pi, 0 \models \phi$. 
If $\pi \models \phi$ for every
$\pi$, then $\phi$ is \emph{true} and we write $\models \phi$.
We abbreviate as $\Diamond \phi$ (``eventually'') the formula $\top$
$\mathcal{U} \phi$ and $\Box \phi$ (``always'') the formula $\neg
\Diamond \neg \phi$. We also consider other Boolean connectives
like ``$\wedge$'' and ``$\rightarrow$'' with the usual meaning. 
Finally, some of the PSPs use the ``weak until'' operator
defined as $p \mathcal{W} q = \Box p \vee (p \mathcal{U} q)$.

\paragraph{LTL satisfiability.} Among various approaches to decide LTL
satisfiability, 
reduction to model checking was proposed in~\cite{rozier2007}
to check the consistency of requirements expressed as LTL formulae.
Given a formula $\phi$ over a set $Prop$ of atomic
propositions, a \emph{universal} model $M$ can
be constructed. Intuitively, a  universal model encodes all the
possible computations over $Prop$ as (infinite) traces, and therefore
$\phi$ is satisfiable precisely when $M$ does not satisfy $\neg \phi$.
In~\cite{rozier2011} a first improvement over this basic strategy is presented
together with the tool PANDA~\cite{panda}, 
whereas in~\cite{li2013} an algorithm based on automata construction
is proposed to enhance performances even further --- the approach is
implemented in a tool called \aalta. Further studies along this
direction include~\cite{li2014} and~\cite{li2013b}. In the latter,
a portfolio LTL satisfiability solver called \polsat\ is proposed to
run different techniques in parallel and return the result of the first
one to terminate successfully.

\begin{figure}\fbox{%
    \begin{minipage}[t]{0.9\textwidth}

      \begin{center}
        \textbf{Response}
        \end{center}
      Describe cause-effect relationships between a pair of
      events/states. An occurrence of the first, the cause, must be
      followed by an occurrence of the second, the effect. Also known
      as Follows and Leads-to.
      
      \par\noindent\rule{\textwidth}{0.4pt}

      \textbf{Structured English Grammar}
  
      \textsl{It is always the case that if P holds, then S eventually holds.}
      
      \par\noindent\rule{\textwidth}{0.4pt}

      \textbf{LTL Mappings}
      
      { 
        \renewcommand{\arraystretch}{1.5}
         \begin{tabular}{l@{\hskip 1cm}l}
         Globally & $\Box (P \rightarrow \Diamond S)$\\
         Before R & $\Diamond R \rightarrow (P \rightarrow
         (\overline{R}\
         \mathcal{U}\ (S \wedge \overline{R})))
         \ \mathcal{U}\ R$\\
         After Q  & $\Box(Q \rightarrow \Box (P \rightarrow \Diamond S))$\\
         Between Q and R & $\Box((Q \wedge \overline{R} \wedge \Diamond R)\rightarrow  (P \rightarrow (\overline{R}\ \mathcal{U}\ (S \wedge \overline{R})))\ \mathcal{U}\ R)$ \\
         After Q until R & $\Box(Q\wedge \overline{R} \rightarrow ((P\rightarrow(\overline{R}\ \mathcal{U}\ (S \wedge\overline{R})))\ \mathcal{W}\ R)$\\
        \end{tabular}
      } 
      \par\noindent\rule{\textwidth}{0.4pt}
      \textbf{Example}
      
      \textsl{If the train is approaching, then the gate shall be closed}.

  \end{minipage}
  }%
  \caption{Response Pattern}
  \label{fig:response}
  
\end{figure}

\paragraph{Property Specification Patterns (PSPs).} The original
proposal of PSPs is to be found in~\cite{dwyer1999}.
They are meant to describe the essential structure of 
system's behaviours and provide expressions of such behaviors in a
range of common formalism. 
An example of a PSP from~\cite{specpatt} is given in
Figure~\ref{fig:response} --- with some part omitted for sake of readability.
A pattern is comprised of a \emph{Name} (Response in
Figure~\ref{fig:response}), an (informal) statement describing the behaviour
captured by the pattern, and a (structured English)
statement~\cite{konrad2005} that should be used to express requirements.
The LTL mappings corresponding to different declinations of the pattern
are also given, where capital letters (P,S,T,R,Q) stands for
Boolean states/events.\footnote{We omitted some aspects which are
  not relevant for our work, e.g., translations to other logics like
  CTL~\cite{dwyer1999}.}
In more detail, a PSP is composed of two parts: ($i$) the
\emph{scope}, and ($ii$) the \emph{body}. 
The \textsl{scope} is the extent of the program execution over which
the pattern must hold, and there are five scopes allowed:
\emph{Globally}, to span the entire scope execution;
\emph{Before}, to span execution up to a state/event;
\emph{After}, to span execution after a state/event;
\emph{Between}, to cover the part of execution from one state/event to
another one; 
\emph{After-until}, where the first part of the pattern continues
even if the second state/event never happens.
For state-delimited scopes, the interval in which the property is
evaluated is closed at the left and open at the right end.
The \textsl{body} of a pattern, describes the behavior that we want to
specify. In~\cite{dwyer1999} the bodies are categorized in 
\emph{occurrence} and \emph{order} patterns.
Occurrence patterns require states/events to occur or
not to occur. Examples of such bodies are \emph{Absence}, where a
given state/event must not occur within a scope, and its opposite
\emph{Existence}. 
Order patterns constrain the order of the
  states/events. Examples of such patterns are \emph{Precedence}, where a
  state/event must always precede another state/event, and
  \emph{Response}, where a state/event must always be followed by
  another state/event within the scope.
Moreover, we included the \emph{Invariant} pattern introduced in~\cite{post2012formalization}, and dictating that  
a state/event must occur whenever another state/event occurs.
Combining scopes and bodies we can construct 55 different types of
patterns. For more details, please visit~\cite{specpatt}.



\paragraph{Related Work.}
In~\cite{lumpe2011} the framework, \textsl{Property
  Specification Pattern Wizard (PSP-Wizard)} is presented, for
machine-assisted definition of temporal formulae capturing
pattern-based system properties.
PSP-Wizard offers a translation into LTL of the patterns encoded in
the tool, but 
it is meant to aid specification, rather than support consistency
checking, and it cannot deal with numerical signals. 
In~\cite{konrad2005}, an extension is presented
to deal with real-time specifications, together with
mappings to Metric temporal logic (MTL), Timed computational tree
logic (TCTL) and Real-time graphical interval logic (RTGIL).
Even if this work is not directly connected with ours, it is worth
mentioning it since their structured English grammar for patterns is
at the basis of our formalism.
The work in~\cite{konrad2005} also provided inspiration to
a recent set of works~\cite{dokhanchi2016,dokhanchi2015} about
a tool, called VI-Spec, to assist the analyst in the elicitation and
debugging of formal specifications.
VI-Spec lets the user specify requirements
through a graphical user interface, translates them to MITL formulae
and then supports debugging of the specification using run-time
verification techniques. 
VI-Spec embodies an approach similar to ours to deal with numerical
signals by translating inequalities to sets of Boolean variables.
However, VI-Spec differs from our work in several aspects, most
notably the fact that it performs debugging rather than consistency,
so the behavior of each signal over time must be known. Also, VI-Spec
handles only inequalities and does not deal with sets of requirements
written using PSPs. 

\section{Constraint Property Specification Patterns}
\label{sec:consistency}

Let us start by defining a \emph{constraint systems} $\mathcal{D}$
as a tuple $\mathcal{D} = (D, R_1, \ldots, R_n, \mathcal{I}$), where
$D$ is a non-empty set called \emph{domain}, and each
$R_i$ is a predicate symbol of arity $a_i$, with   
$\mathcal{I}$($R_i)$ $\subseteq$ $D^{a_i}$ being its
interpretation. An (atomic) $\mathcal{D}$-\emph{constraint} over a set of
variables $X$  is of the form $R_i(x_1, \ldots, x_{a_i})$ for some
$1 \leq i \leq n$ and $x_j \in X$ for all $1 \leq j \leq a_i$ --- we
also use the term   
\emph{constraint} when $\mathcal{D}$ is understood from the context.
We define \emph{linear temporal logic modulo constraints}
--- LTL($\mathcal{D}$) for short --- as an extension of LTL with atoms
in a constraint system $\mathcal{D}$. Given a set of Boolean 
propositions $Prop$, a constraint system $\mathcal{D} = (D, R_1,
\ldots, R_n, \mathcal{I})$, and a set of variables $X$, an
LTL($\mathcal{D}$) formula is defined as: 
\begin{center}
  $\phi =$ $\bot$ $\mid$ $\top$ $\mid$ $p$ $\mid$ $R_i(x_1, \ldots,
  x_{a_i})$ $\mid$ $\neg \phi_1$ 
  $\mid$ $\phi_1\vee\phi_2$ $\mid$ $\mathcal{X}
  \phi_1$ $\mid$ $\phi_1 \mathcal{U} \phi_2$ 
\end{center}
where $p \in Prop$, $\phi, \phi_1, \phi_2$ are LTL($\mathcal{D}$)
formulas,  and $R_i(\cdot)$ with $1 \leq i \leq n$ is an atomic
$\mathcal{D}$-constraint. Additional Boolean and temporal operators
are defined as in LTL with the same intended
meaning. Notice that the set of LTL($\mathcal{D}$) formulas is a
(strict) subset of those in constraint linear temporal logic ---
CLTL($\mathcal{D}$) for short --- as defined, e.g.,
in~\cite{demri2002}. LTL($\mathcal{D}$) formulas are also interpreted 
over computations of the form $\pi: \mathbb{N} \rightarrow 2^{Prop}$
plus additional \emph{evaluations} of the form $\nu: X \times
\mathbb{N} \rightarrow D$, i.e., $\nu$ is a function assigning at
each variable $x \in X$ a corresponding value $\nu(x,i)$
at each time instant $i \in \mathbb{N}$. LTL semantics is extended to
LTL($\mathcal{D}$) by handling constraints:
\begin{center}
  $\pi,\nu,j \models R_i(x_1, \ldots, x_{a_i})$ iff
  $(\nu(x_1,j), \ldots, \nu(x_{a_i},j)) \in \mathcal{I}(R_i)$
\end{center}
We say that $\pi$ and $\nu$ \emph{satisfy} a formula $\phi$, denoted
$\pi, \nu \models \phi$, iff $\pi, \nu, 0 \models \phi$. A formula
$\phi$ is \emph{satisfiable} as long as there exist a computation
$\pi$ and a valuation $\nu$ such that $\pi, \nu \models \phi$. 
We further restrict our attention to the constraint system $D_C$ =
($\mathbb{R}$,$<$,$=$), with atomic constraints of the form $x < C$
and $x = C$, where $C \in \mathbb{R}$ is a constant.
While CLTL($\mathcal{D}$) is undecidable in
general~\cite{demri2002,comon2000}, LTL$(\mathcal{D}_C)$ is decidable
since, as we show in the following, it can be reduced to LTL satisfiability.    

We introduce the concept of \emph{constraint property specification
  pattern}, denoted PSP($\mathcal{D}$), to deal with specifications
containing Boolean variables as well as atoms from a constraint system 
$\mathcal{D}$. In particular, a PSP($\mathcal{D}_C$) features only
Boolean atoms and atomic constraints of the form
$x < C$ or $x = C$ ($C \in \mathbb{R}$). For example, the requirement:
\begin{center}
  \emph{The angle of joint1 shall never be greater than 170 degrees}
\end{center}
can be re-written as a PSP($\mathcal{D}_C$):
\begin{center}
  \emph{Globally, it is always the case that $\theta_1 < 170$} 
\end{center}
where $\theta_1 \in \mathbb{R}$ is the variable associated to the
angle of \emph{joint1} and $170$ is the limiting threshold.
While basic PSPs only allow for Boolean states/events in
their description, PSPs($\mathcal{D}_C$) also allow for atomic
constraints. It is straightforward to extend the translation
of~\cite{dwyer1999} from basic PSPs to LTL in order to encode any
PSP($\mathcal{D}_C$) to a formula in LTL($\mathcal{D}_C$). Consider,
for instance, the set of requirements:
\begin{itemize}
\item[$R_1$] Globally, it is always the case that \textbf{v $\leq$
  5.0} holds.
\item[$R_2$] After \textbf{a}, \textbf{v $\leq$ 8.5} eventually holds.
\item[$R_3$] After \textbf{a}, it is always the case that if
  \textbf{v $\geq$ 3.2} holds, then \textbf{z} eventually holds.
\end{itemize}
where \textbf{a} and \textbf{z} are Boolean states/events, whereas
\textbf{v} is a numeric signal. 
These PSPs($\mathcal{D}_C$)\footnote{Strictly speaking, the syntax
  used is not that of $\mathcal{D}_C$, but a statement like $v \leq
  5.0$ can be thought as syntactic sugar for the expression $(v <
  5.0) \vee (v = 5.0)$.}
can be rewritten as the following LTL($\mathcal{D}_C$) formula: 
\begin{equation}
  \label{eq:reqex}
  \begin{array}{ll}
    \Box (v < 5.0 \vee v = 5.0) & \wedge \\
    \Box (a \rightarrow \Diamond (v < 8.5) \vee (v = 8.5)) & \wedge \\
    \Box (a \rightarrow \Box (\neg (v < 3.2) \rightarrow \Diamond z))
  \end{array}
\end{equation}
Therefore, to reason about the consistency of sets of requirements
written using PSPs($\mathcal{D}_C$) it is sufficient to provide an
algorithm for deciding the satisfiability of LTL($\mathcal{D}_C$)
formulas.

To this end, consider an LTL($\mathcal{D}_C$) formula $\phi$, and 
let $X(\phi)$ be the set of variables that occur in $\phi$. 
We define the \emph{set of thresholds} $T_x(\phi)$ as the set of
constant values against which variable $x \in X(\phi)$ is compared to.
More precisely, for every variable $x \in X(\phi)$ we construct a
set $T_x(\phi)$ = $\{t_1,..,t_n\}$ such that, for all $t_i \in
\mathbb{R}$ with $1 \leq i \leq n$, $\phi$ contains a constraint of
the form $x < t_i$ or $x = t_i$. In the following, for our
convenience, we consider each threshold set $T_x(\phi)$ ordered in
ascending order, i.e., $t_{i} < t_{i+1}$ for all $1 \leq i < n$. For
instance, in example (\ref{eq:reqex}), we have $X = \{v \}$ and the set
$T_v = \{ 3.2, 5.0, 8.5 \}$. Given an LTL($\mathcal{D}$) formula
$\phi$, let $T_x(\phi) = \{t_1, \ldots, t_n \}$ be the ordered set of
thresholds for some variable $x \in X(\phi)$; given a computation
$\pi$ and a valuation $\nu$ we can define:
\begin{itemize}
\item $C_x(\phi)$ as the set of Boolean variables such that for each
  $c_j \in C_x(\phi)$ we have $c_j \in \pi(i)$ for $i = 0, 1, \ldots$
  exactly when $t_{j-1} < \nu(x,i) < t_j$, if $j >1$, and $\nu(x,i) <
  t_j$, if $j = 1$  with $t_j \in T_x(\phi)$ for all $1 \leq j \leq n$.
\item $E_x(\phi)$ as the set of Boolean variables such that
  for each $e_j \in E_x(\phi)$ we have $e_j \in \pi(i)$ for $i = 0, 1,
  \ldots$ exactly when $\nu(x,i) = t_j$ for some $t_j \in T_x(\phi)$.
\end{itemize}
Notice that, by definition of $C_x(\phi)$ and $E_x(\phi$), given any
time instant $i \in 0, 1, 2, \ldots$, we have that exactly
one of the following cases is true ($1 \leq j \leq n$):
\begin{itemize}
  \item $c_j \in \pi(i)$ for some $j$, $c_l \not\in
    \pi(i)$ for all $l \neq j$ and  $e_j \not\in \pi(i)$ for all $j$;
  \item $e_j \in \pi(i)$ for some $j$, $e_l \not\in
    \pi(i)$ for all $l \neq j$ and  $c_j \not\in \pi(i)$ for all $j$;
  \item $c_j \not\in \pi(i)$ and $e_j \not\in \pi(i)$ for all $j$.
\end{itemize}
Intuitively, the first case above corresponds to a value
of $x$ that lies between some threshold value in
$T_x(\phi)$ or before its smallest value; the second case occurs
when a threshold value is assigned to $x$, and the third case is when
$x$ exceeds the highest threshold value in $T_x(\phi)$.
For instance, in example (\ref{eq:reqex}) we have
$T_v = \{ 3.2,  5.0, 8.5 \}$ and the corresponding sets
$C_v \{c_1, c_2, c_3 \}$ and $E_v = \{e_1, e_2 ,e_3\}$. Assuming,
e.g., $\nu(v,i) = 10$ for some $i = 0, 1, 2, \ldots$, we would have
that $C_v \cap \pi(i) = E_v \cap \pi(i) = \emptyset$.

Given the definitions above, an LTL($\mathcal{D}$) formula $\phi$ over the
set of Boolean propositions $Prop$ and the set of variables $X$, can be
converted to an LTL formula $\phi'$ over the set of Boolean
propositions $Prop \cup \bigcup_{\xi in X} (C_\xi(\phi) \cup E_\xi(\phi))$ using the following
substitutions: 
\begin{equation}
  \label{eq:encoding}
  x < t_i \leadsto \bigvee_{j=1}^i c_j \vee \bigvee_{j=1}^{i-1}e_j
  \qquad \mbox{and} \qquad
    x = t_i \leadsto e_j.
\end{equation}
However, replacing atomic constraints is not enough to ensure
equisatisfiability of $\phi'$ with respect to $\phi$. In particular, 
we must encode the observation made above about ``mutually exclusive'' 
Boolean valuations for variables in $C_x(\phi)$ and $E_x(\phi)$ for
every $x \in X(\phi)$ as corresponding Boolean constraints:
\begin{equation}
  \label{eq:phiM}
    \phi_M = \bigwedge_{\xi \in X(\phi)} \left( \bigwedge_{a,b \in M_{\xi(\phi)}, a \neq
    b} \Box \neg (a \wedge b) \right) 
\end{equation}
where $M_\xi(\phi) = C_\xi(\phi) \cup E_\xi(\phi)$. We can now state
the following fact:
\begin{property}
  \label{prop:key}
Given an LTL($\mathcal{D}_C$) formula $\phi$ over the set of Boolean
atoms $Prop$ and variables $X(\phi)$, and the corresponding sets
$C_x(\phi)$ and $E_x(\phi)$ defined for all $x \in X(\phi)$ as
described above, we have that $\phi$ is satisfiable if and only if the
LTL formula $\phi_M \rightarrow \phi'$ is satisfiable, where $\phi'$ is 
obtained by replacing atomic constraints according to rules
(\ref{eq:encoding}) and $\phi_M$ is defined according to (\ref{eq:phiM}).
\end{property}
For instance, given example (\ref{eq:reqex}), we have $C_v = \{c_1,
c_2, c_3\}$ and $E_v = \{e_1, e_2, e_3\}$ and the mutual exclusion constraints
are written as: 
\begin{equation}
\begin{split}
  \phi_{M} =&
  \Box \neg (c_1 \wedge c_2) \wedge \Box \neg (c_1 \wedge c_3) \wedge 
  \Box \neg (c_1 \wedge e_1) \wedge \Box \neg (c_1 \wedge e_2) \wedge  \\&
  \Box \neg (c_1 \wedge e_3) \wedge \Box \neg (c_2 \wedge c_3) \wedge
  \Box \neg (c_2 \wedge e_1) \wedge \Box \neg (c_2 \wedge e_2) \wedge  \\&
  \Box \neg (c_2 \wedge e_3) \wedge \Box \neg (c_3 \wedge e_1) \wedge
  \Box \neg (c_3 \wedge e_2) \wedge \Box \neg (c_3 \wedge e_3) \wedge  \\&
  \Box \neg (e_1 \wedge e_2) \wedge \Box \neg (e_1 \wedge e_3) \wedge
  \Box \neg (e_2 \wedge e_3). 
\end{split}
\end{equation}
Therefore, the LTL formula to be tested for assessing the consistency of
the requirements is
\begin{equation}
  \label{eq:reqex2}
  \begin{array}{ll}
    \phi_M \rightarrow ( & \Box (c_1 \vee c_2 \vee e_1 \vee e_2) \wedge \\
                         & \Box (a \rightarrow \Diamond (\bigvee_{i=1}^3 c_i \vee e_i)) \wedge \\
                         & \Box (a \rightarrow \Box (\neg (c_1 \vee e_1) \rightarrow \Diamond z))).
  \end{array}
\end{equation}

\section{Analysis with Probabilistic Requirement Generation}
\label{sec:random}

The main goal of this Section is to investigate the scalability of our
encoding from LTL($\mathcal{D}$) to LTL.  To this end, we evaluate the
performances\footnote{All the experiments reported in this Section ran
  on a server equipped with 2 Intel Xeon E5-2640 v4 CPUs and 256GB RAM
  running Debian with kernel 3.16.0-4.} of some
state-of-the-art tools for LTL satisfiability, and then we consider
the best among such tools to assess whether our approach can scale to
sets of requirements of realistic size. Since we want to have
control over the kind of requirements, as well as the number of
constraints and the size of the corresponding domains, we generate
artificial specifications using a probabilistic model that we devised
and implemented specifically to carry out the experiments herein
presented.
In particular, the following parameters can be tuned in our
generator of specifications:
\begin{itemize}
\item The number of requirements generated ($\#req$).
\item The probability of each different body to occur in a
  pattern. 
\item The probability of each different scope to occur in a
  pattern. 
\item The size ($\#vars$) of the set from which variables
  are picked uniformly at random to build patterns.
\item The size ($dom$) of the domain from which the thresholds of the
  atomic constraints are chosen uniformly at random.
\end{itemize}

\paragraph{Evaluation of LTL satisfiability solvers.}
The solvers considered in our analysis are the ones included in
the portfolio solver \polsat~\cite{li2013b}, namely
\aalta~\cite{li2013}, \nusmv~\cite{cimatti2002},
\pltl~\cite{schwendimann1998}, and \trp~\cite{hustadt2003}.  In order
to have a better understanding about the behavior of such solvers,
we ran them separately instead of running \polsat. Furthermore,
in the case of \nusmv, 
we considered two different encodings. With reference to
Property~\ref{prop:key}, the first encoding defines $\phi_M$ 
as an invariant --- denoted as \nusmvi{} --- and $\phi'$ is the property
to check; the second encoding considers $\phi_M \rightarrow \phi$ as
the property to check --- denoted as \nusmvn{}.
In our experimental analysis we set the range of the parameters
as follows: $\#vars$ $\in \{16,32\}$, $dom \in \{2,4,8,16\}$, and
$\#req$ $\in\{8, 16, 32,64\}$.  For each combination of the
parameters with $v\in \#vars$, $r\in \#req$ and $d\in dom$, we generate
10 different benchmarks. Each benchmark is a specification
containing $r$ requirements where each scope has
(uniform) probability 0.2 and each body has (uniform)
probability 0.1. Then, for each atomic constraint in 
the benchmark, we choose a variable out of $v$ possible ones, and a
threshold value out of $d$ possible ones.
In Table~\ref{tab:preliminary} we show the results of the
analysis.  Notice that we do not show the results of \trp\ because of
the high number of failures obtained. Looking at the table, we can see
that \aalta\ is the tool with the best performances, as it is capable
of solving two times the problems solved by other solvers in most
cases. Moreover, \aalta\ is up to 3 orders of magnitude faster than
its competitors. Considering unsolved instances, it is worth noticing
that in our experiments \aalta\ never reaches the granted time limit (10
CPU minutes), but it always fails beforehand. This is probably due to
the fact that \aalta\ is still in a relatively early stage of
development and it is not as mature as \nusmv\ and \pltl. Most
importantly, we did not found any discrepancies in the satisfiability
results of the evaluated tools. 
\begin{table}[t!]
\setlength{\tabcolsep}{2pt}
\caption{Evaluation of LTL satisfiability solvers on
  randomly generated requirements. The first line reports the size of
  the domain ($dom$), while the second line reports the total amount
  of variables ($vars$) for each domain size. Then, for each tool (on
  the first column), the table shows the total amount of solved
  problems and the CPU time (in seconds) spent to solve them (columns
  ``S'' and ``T'', respectively).}
\label{tab:preliminary}
{\scriptsize
\begin{tabular}{|l|r|r|r|r|r|r|r|r|r|r|r|r|r|r|r|r|}\hline
\multicolumn{1}{|c|}{$dom$} & \multicolumn{4}{|c|}{2} & \multicolumn{4}{|c|}{4} & \multicolumn{4}{|c|}{8} & \multicolumn{4}{|c|}{16} \\\hline
  \multicolumn{1}{|c|}{$\#vars$} & \multicolumn{2}{|c|}{16} & \multicolumn{2}{|c|}{32} & \multicolumn{2}{|c|}{16} & \multicolumn{2}{|c|}{32} & \multicolumn{2}{|c|}{16} & \multicolumn{2}{|c|}{32} & \multicolumn{2}{|c|}{16} & \multicolumn{2}{|c|}{32} \\\hline
\multicolumn{1}{|c|}{\bf Tool} & \multicolumn{1}{|c|}{\bf S} & \multicolumn{1}{|c|}{\bf T} & \multicolumn{1}{|c|}{\bf S} & \multicolumn{1}{|c|}{\bf T} & \multicolumn{1}{|c|}{\bf S} & \multicolumn{1}{|c|}{\bf T} & \multicolumn{1}{|c|}{\bf S} & \multicolumn{1}{|c|}{\bf T} & \multicolumn{1}{|c|}{\bf S} & \multicolumn{1}{|c|}{\bf T} & \multicolumn{1}{|c|}{\bf S} & \multicolumn{1}{|c|}{\bf T} & \multicolumn{1}{|c|}{\bf S} & \multicolumn{1}{|c|}{\bf T} & \multicolumn{1}{|c|}{\bf S} & \multicolumn{1}{|c|}{\bf T} \\ \hline \hline
  \aalta & \textbf{16} & 0.0 & \textbf{27} & 0.1 & \textbf{22} & 0.1 & \textbf{29} & 0.4 & \textbf{26} & 0.6 & \textbf{29} & 1.4 & \textbf{25} & 2.8 & \textbf{31} & 4.9 \\ \hline
  \nusmvi   & 11 & 30.4 & 10 & 185.1 & 10 & 804.2 & 9 & 881.3 & 11 & 68.1 & 8 & 402.9 & 10 & 1172.6 & 8 & 1001.9\\ \hline
  \nusmvn & 11 & 65.0 & 10 & 489.7 & 7 & 303.6 & 7 & 505.5 & 11 & 92.4 & 10 & 1277.6 & 8 & 660.0 & 9 & 1394.5 \\ \hline
  \pltl        & 8 & 25.0 & 11 & 108.1 & 9 & 1.2 & 10 & 0.6 & 10 & 19.6 & 11 & 0.1 & 11 & 14.5 & 14 & 3.5 \\ \hline
\end{tabular}
}
\end{table}
\begin{figure}[p]
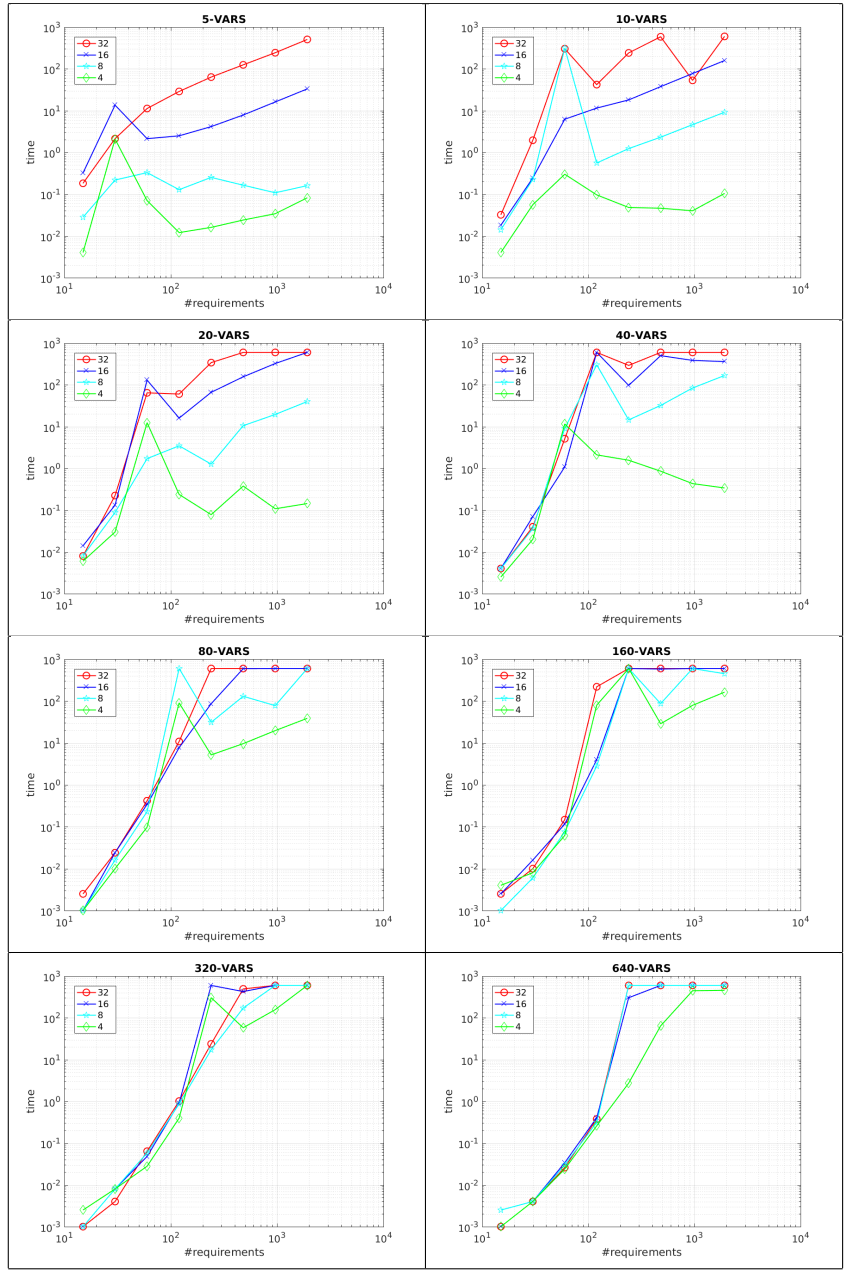

\begin{center}
\begin{tabular}{|l|l|}\hline
  \pgfuseimage{i1} & \pgfuseimage{i2} \\\hline
  \pgfuseimage{i3} & \pgfuseimage{i4} \\\hline
  \pgfuseimage{i5} & \pgfuseimage{i6} \\\hline
  \pgfuseimage{i7} & \pgfuseimage{i8} \\\hline
\end{tabular}
\end{center}
\caption{\label{fig:random} Scalability Analysis. On the $x$-axes
  ($y$-axes resp.) we report $\#req$ (CPU time in seconds resp.). Axis
  are both in logarithmic scale. In each plot we consider different
  values of $\#dom$. In particular, the diamond green line is for
  $\#dom$ = 4, the light blue line with stars is for $\#dom$ = 8, the
  blue crossed lines and red circled ones denote $\#dom = 16$ and
  $\#dom = 32$, respectively.}
\end{figure}
\paragraph{Evaluation of scalability.}
The analysis involves 2560 different benchmarks generated as
in the previous experiment.  The initial
value of $\#req$ has been set to 15, and it has been doubled until
1920, thus obtaining benchmarks with a total amount of requirements
equals to 15, 30, 60, 120, 240, 480, 960, and 1920. Similarly has been
done for $\#vars$ and $\#dom$; the former ranges from 5 to 640, while
the latter ranges from 4 to 32.  At the end of the generation, we
obtained 10 different sets composed of 256 benchmarks. 
In Figure~\ref{fig:random} we present the results,
obtained running \aalta. The Figure is composed by 8
plots, one for each value of $\#vars$. 
Looking at the plots in Figure~\ref{fig:random}, we can see that
the difficulty of the problem increases when all the values
of the considered parameters increase, and this is particularly true
considering the total amount of requirements. The parameter
$\#dom$ has a higher impact of difficulty when the number of
variables is small. Indeed, when the number of variables is less then 40
there is a clear difference between solving time with $\#dom = 4$
and $\#dom = 32$.  On the other hand when the number of variables
increases, all the plots for various values of $\#dom$ are very close
to each other. As a final remark, we can see that even considering the
largest problem ($\#vars$ = 640, $\#dom$ = 32), more than the 60\% of
the problems are solved by \aalta\ within the time limit of 10 minutes.

\begin{wrapfigure}{l}{0.5\textwidth}
  \begin{center}
    \includegraphics[scale=0.3]{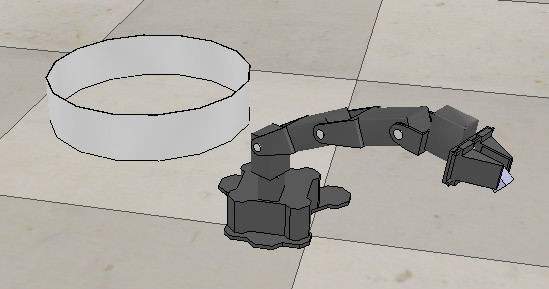}
  \end{center}
  \caption{\label{fig:robot}WidowX robotic arm moving a grabbed
    object in the bucket on the left.}
\end{wrapfigure}

\section{Analysis with a Controller for a Robotic Manipulator}
\label{sec:robot}

In this Section, as a basis for our experimental analysis, we
consider a set of requirements from the design of an embedded
controller for a robotic manipulator. The controller should direct a
properly initialized robotic arm --- and related vision system --- to
look for an object placed in a given position and move to such
position in order to grab the object; once grabbed, the object is to
be moved into a bucket placed in a given position and released without
touching the bucket. The robot must stop even in the case of an
unintended collision with other objects or with the robot itself ---
collisions can be detected using torque estimation from current
sensors placed in the joints. Finally, if a general alarm 
is detected, e.g., by the interaction with a human supervisor, the
robot must stop as soon as possible.
The manipulator is a 4 degrees-of-freedom Trossen Robotics WidowX
arm\footnote{Technical specifications are available at
  \url{http://www.trossenrobotics.com/widowxrobotarm}.} equipped
with a gripper: Figure~\ref{fig:robot} shows a snapshot of the robot
in the intended usage scenario taken from
V-REP\footnote{\url{http://www.coppeliarobotics.com/}} simulator.
The design of the embedded controller is currently
part of the activities related to the ``Self-Healing System for
Planetary Exploration'' use case~\cite{masin2017cross} in the context
of the EU project CERBERO.


\begin{table}[t!]
  \begin{tabular}{|l|r|r|r|r|r|r|} \hline
    \multicolumn{1}{|c|}{\textbf{Pattern}} &
    \multicolumn{3}{|c|}{Specification} & \multicolumn{3}{|c|}{Fault
      injections} \\ \cline{2-7}
    \multicolumn{1}{|c|}{\ } & \multicolumn{1}{|c|}{\textsc{after}} &
    \multicolumn{1}{|c|}{\textsc{after\_until}} &
    \multicolumn{1}{|c|}{\textsc{globally}} &
     \multicolumn{1}{|c|}{\textsc{after}} &
     \multicolumn{1}{|c|}{\textsc{after\_until}} &
     \multicolumn{1}{|c|}{\textsc{globally}} \\ \hline\hline
    Absence  & -- & 12 & 14 & [F4] & -- & [F3] \\ \hline
    Existence & 9 & -- & -- & -- & [F5] & [F4, F6] \\ \hline
    Invariant & -- & -- & 29 & -- & -- & [F2, F6]\\ \hline
    Precedence  & -- & -- & 1 & -- & -- & -- \\ \hline
    ResponseChain & -- & -- & 2 & -- & -- & -- \\ \hline
    Response & 1 & -- & 4 & -- & -- & [F1] \\ \hline
    Universality & 2 & -- & 1 & -- & -- & -- \\ \hline 
  \end{tabular}
  \caption{  \label{tab:robot-req}%
    Robotic use case requirements synopsis. The table is
    organized as follows: the first column reports the name of
    the patterns and it is followed by two groups of three columns
    denoted with the scope type: the first group refers to the intended
    specification, the second to the one with fault injections.
    Each cell in the first 
    group reports the number of requirements grouped by pattern and by
    scope type. Cells in the second group categorize the 6 injected
    faults, labeled with F1, \ldots, F6. 
  }
\end{table}

In this case study, constrained numerical signals are used to
represent requirements related to various parameters, namely
angle, speed, acceleration, and torque of the 4 joints,
size of the object picked, and force exerted by the end-effector. 
We consider 75 requirements, including those involving
scenario-independent constraints like joints limits, 
and mutual exclusion among states, as well as specific requirements
related to the conditions to be met at each state. The set of
requirements involved in our analysis includes 14 Boolean signals and
20 numerical ones. The full list of requirements is available
at~\cite{robotArmUsecase}, each one of them expressed as a
PSP($\mathcal{D}$). In Table~\ref{tab:robot-req} we present a synopsis
  of the requirements, to give an idea of the kind of patterns used in
  the specification. 

Our first experiment\footnote{Experiments
herein presented ran on a PC equipped with a CPU Intel Core i7-2760QM @
2.40GHz (8 cores) and 8GB of RAM, running Ubuntu 14.04 LTS.}
is to run \nusmvi{} on the intended specification translated to
LTL($\mathcal{D}_C$). The motivation for presenting the results
with \nusmvi{} rather than \aalta{} is twofold: While its performances
are worse than \aalta{}, \nusmvi{} is more robust in the sense that it
either reaches the time limit or it solves the problem, without ever
failing for unspecified reasons like \aalta{} does at times; second, it
turns out that \nusmvi{} can deal flawlessly and in reasonable CPU times
with all the 
specifications we consider in this Section, both the intended one and
the ones obtained by injecting faults. In particular, on the intended
specification, \nusmvi{} is able to find a counterexample in 37.1 CPU
seconds, meaning that there exists at least a model able to    
satisfy all the requirements simultaneously. Notice that the
translation time from patterns to formulas in LTL($\mathcal{D}_C$) is
negligible with respect to the solving time. 
Our second experiment is to run \nusmvi{} on the specification with
some faults injected. In particular, we consider six different
faults, and we extend the specification in six different ways
considering one fault at a time. The patterns related to the faults
are summarized in Table~\ref{tab:robot-req}. Also in this case, we
refer the reader to~\cite{robotArmUsecase} for details.
In case of faulty specifications, \nusmvi{} concludes that no
counterexample exists, i..e, there is no model able to satisfy all the
requirements simultaneously.  In particular, in the case of 
F2 and F3, \nusmvi{} returned the result in 2.1 and 1.7 CPU seconds,
respectively. Concerning the other faults, the tools was one order of 
magnitude slower in returning the satisfiability result. In 
particular, it spent 16.8, 50.4, 12.2, and 25.6 CPU seconds in the
evaluation of the requirements when faults 1, 4, 5 and 6 are injected,
respectively.

The noticeable difference in performances when checking for different
faults in the specification is mainly due to the fact that F2 and F3
introduce an initial inconsistency, i.e., it would not be possible to
initialize the system if they were present in the specification,
whereas the remaining faults introduce inconsistencies related to
interplay among constrains in time, and  thus additional search is
needed to spot problems. In order to explain this difference, let
us first consider fault 2: 
\begin{center}
  \parbox{0.7\textwidth}{
  \emph{Globally, it is always the case that if} \texttt{state\_init}
  \emph{holds, then not } \texttt{arm\_idle} \emph{ holds as well.}
  }
\end{center}
It turns out that in the intended specification there
is one requirement specifying exactly the opposite, i.e.,
that when the robot is in \texttt{state\_init}, then \texttt{arm\_idle}
must hold as well. Thus, the only models that satisfy both requirements
are the ones preventing the robot arm to be in
\texttt{state\_init}. However, this is not possible because other 
requirements related to the state evolution of the system impose that
\texttt{state\_init} will eventually occur and, in particular, that it
should be the first one.
On the other hand, if we consider fault 6:
\begin{center}
  \parbox{0.7\textwidth}{
{\it Globally, it is always the case that if} \texttt{arm\_moving}
{\it holds, then} \texttt{joint1\_speed} $>$ {\it 15.5 holds as
  well.}\\ {\it Globally,} \texttt{arm\_moving} {\it and}
\texttt{proximity\_sensor} {\it = 10.0 eventually holds.}
}
\end{center}
we can see that the first requirement sets a lower speed bound at 15.5
$deg/s$ for \texttt{joint1} when the arm is moving, while there
exists a requirement in the intended specification setting an upper
speed bound at 10 $deg/s$ when the proximity sensor detects an object
closer than 20 $cm$. In this case, the model checker is still able to
find a valid model in which \texttt{proximity\_sensor} $<$ 20.0 never
happens when \texttt{arm\_moving} holds, but the second requirements
in fault 6 prohibits this opportunity. It is exactly this kind of
interplay among different temporal properties which makes \nusmvi{}
slower in assessing the (in)consistency of some specifications.


\section{Conclusions}
\label{sec:concl}
Enabling the verification of high-level requirements is one of the key
aspects towards the development of safety- and security-critical
cyber-physical systems.
Property Specification Patterns offer a viable path towards
this target, but their expressiveness is often too restricted for
practical applications.
In this paper, we have extended basic PSPs over the constraint system
$\mathcal{D}_C$, and we have provided an encoding from any
PSP($\mathcal{D}_C$) into a corresponding LTL formula.
This enables us to deal with many specifications of practical interest,
and to verify them using  automated reasoning systems currently
available for LTL. 
Using realistically-sized specifications generated with an artificial
probability model we have shown that our approach implemented on the
tool \aalta{} scales to problems containing more than a thousand
requirements over hundreds of variables. Considering a real-world case
study in the context of the EU project CERBERO, we have shown
that it is feasible to check specifications and uncover injected
faults, even without resorting to \aalta{}, but considering the
(slower, yet more robust) \nusmv. These results witness that our
approach is viable and worth of adoption in the process of requirement
engineering.  Our next steps toward this goal will include easing the
translation from natural language requirements to patterns, and
extending the pattern language to deal with other relevant aspects of
cyber-physical systems, e.g., real-time constraints.

\paragraph{Acknowledgments} The research of Luca Pulina and Simone Vuotto has been funded by the EU Commission's H2020 Programme under grant agreement N. 732105 (CERBERO project).

\bibliographystyle{splncs03} 



\end{document}